\documentclass[aps,prl,showpacs,twocolumn,groupedaddress]{revtex4}

\usepackage{amsmath, amsfonts, amsthm, amssymb, graphicx}
\def\op{\ensuremath{\mathcal O}}

\begin{document}


\title{Real-time gauge/gravity duality}


\author{Kostas Skenderis}
\email[]{skenderi@science.uva.nl}
\author{Balt C. van Rees}
\email[]{brees@science.uva.nl}
\affiliation{Institute for Theoretical Physics, Valckenierstraat 65,
1018 XE Amsterdam}


\date{\today}

\begin{abstract}
We present a general prescription for the holographic computation of
real-time $n$-point functions in non-trivial states. In QFT such
real-time computations involve a choice of a time contour in the
complex time plane. The holographic prescription amounts to
``filling in'' this contour with bulk solutions: real segments of
the contour are filled in with Lorentzian solutions while imaginary
segments are filled in with Riemannian solutions and appropriate
matching conditions are imposed at the corners of the contour. We
illustrate the general discussion by computing the 2-point function
of a scalar operator using this prescription and by showing that this
leads to an unambiguous answer with the correct $i \epsilon$
insertions.
\end{abstract}


\maketitle

The gravity/gauge theory duality has been one of the most far
reaching developments in recent years. On the one hand it
opens a window into strong coupling dynamics of gauge theories and
on the other hand it provides a realization of holography and offers
a new perspective in gravitational physics. In recent times,
it has found applications that range from phenomenology to condensed matter
physics.

The foundational papers on the subject
\cite{Maldacena:1997re} laid down
the  basic principles of the duality. The detailed
dictionary between bulk and boundary physics, however,
is best understood to date in the supergravity approximation
and in the Euclidean regime,
i.e. the bulk solution involves a hyperbolic
Riemannian manifold and the boundary theory is Wick-rotated.
While this suffices
for many applications, there are also many reasons for developing a
general real-time prescription. Such a real-time formalism should be used,
for example, in studies of time-dependent phenomena, analysis of
gauge theories in nontrivial pure or mixed states, or
the holographic interpretation of non-stationary spacetimes.

Such a formalism, applicable
at the same level of generality as the corresponding Euclidean
prescription, would constitute an integral part of the definition
of the holographic correspondence and as such is important on
general grounds. Furthermore, there is
an urgency for setting up such a formalism since
interesting current applications, for example
the holographic modelling of the quark-gluon plasma, crucially
involve real-time physics. Actually much of the recent work
on real-time holographic prescriptions was driven by such
applications, see \cite{Son:2007vk} for a review. The aim of this
work is to provide a concrete, first principles prescription
that covers all $n$-point functions and is applicable for
any QFT that has (an Asymptotically AdS) holographic dual. Previous
work on this subject includes \cite{Herzog:2002pc,Son:2002sd,Marolf:2004fy}
and our results agree with these works when we restrict to
their respective domains of validity.

The basic Euclidean holographic dictionary identifies the
boundary conditions $\phi_{(0)}$ for the bulk fields $\Phi$
to sources of the dual
boundary operators and the bulk partition function, which is a
functional of these boundary fields, to the generating
functional of connected $n$-point functions. The main new issue that
arises in the Lorentzian context is that in the bulk, on top of specifying
boundary conditions $\phi_{(0)}$, one also needs to specify initial and/or
final conditions $\phi_{\pm}$ for all fields,
and the bulk partition function is also 
a functional of these. The main question is to understand their meaning
in the dual QFT.
Intuitively, $\phi_\pm$ should be related to QFT in- and
out-states \cite{Balasubramanian:1998sn}, but an exact
prescription to translate QFT states to initial and final boundary
data for the bulk fields has not previously been worked out.

\begin{figure}
\includegraphics[width=5cm]{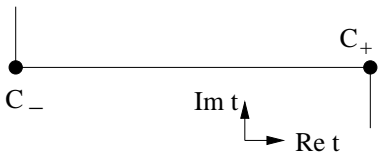}
\caption{\label{fig:contour}The contour in the complex time plane used to obtain vacuum-vacuum amplitudes.}
\end{figure}

Let us briefly recall some QFT basics
that are relevant to our discussion. Consider
a field configuration with initial condition $\phi_-(\vec{x})$ at
$t=-T$ and final condition $\phi_+(\vec{x})$ at $t=T$. The path integral
with fields constrained to satisfy these conditions produces the
transition amplitude $\langle \phi_+,T|\phi_-, -T\rangle$.
If we are interested in vacuum amplitudes we should multiply this
expression by the vacuum wavefunction $\langle 0 |\phi_+, T\rangle$
and  $\langle \phi_-,-T| 0\rangle$ and integrate over $\phi_+,\phi_-$.
The insertion of these wave functions is equivalent to extending the
fields in the path integral to live along a contour in the complex time
plane as sketched in Fig.~\ref{fig:contour}. Indeed, the infinite
vertical segment starting at $-T$ corresponds to a transition amplitude
$\lim_{\beta \to \infty} \langle \phi_-,-T | e^{-\beta H} | \Psi \rangle$
for some state $|\Psi\rangle$, which is however irrelevant since taking the
limit projects it onto the vacuum wave function $\langle \phi_-,-T | 0\rangle$. Similarly, we obtain $\langle 0 | \phi_+, T\rangle$ from the vertical segment starting at $t=T$.
Recall also that these wave function insertions
ultimately lead to the $i \epsilon$ factors in the Feynman propagators.

If one wants to compute expectation values in
non-trivial states or thermal ensembles then one should consider different time
contours, like a real-time thermal contour or a closed in-in contour.
\\

\paragraph{Prescription.} The holographic prescription we propose is to use
``piece-wise'' holography: for each real segment of the time
contour  we consider a Lorentzian solution and for each imaginary
part an Euclidean solution. At the corners of the contour, the various bulk solutions
are joined together using standard matching conditions, i.e. the induced values of the
fields and their conjugate momenta should be (appropriately) continuous
along the gluing surface, which is some hypersurface in the bulk. This results in a completely holographic prescription where all data are encoded in the conformal boundary of
the entire spacetime; the initial and final states are encoded in
the boundary of the Euclidean parts.

The next step is to compute the value of the combined
(Euclidean plus Lorentzian) on-shell actions and then vary these w.r.t. sources to obtain the
renormalized holographic 1-point functions in the presence of sources.
This results in a formula that relates the
1-point function to the asymptotics of the bulk solution \cite{de Haro:2000xn}. Recall that the
holographic renormalization required in this procedure relies on having
sufficient control over the asymptotics of the bulk solutions.
This analysis, however,
is independent of the signature of the spacetime, so all
standard results carry over immediately. One only needs to check that the
corners where Lorentzian and Euclidean solutions are joined do not introduce
any complications. The matching conditions ensure that
this is the case, as will be described elsewhere \cite{SvR}.

As usual, to compute holographic $n$-point functions we need to solve
the bulk field equations to order $(n-1)$ around the
bulk solution. The result should then be
substituted in the $(n-1)$th variation of the holographic 1-point
function to obtain the $n$-point function.
Of course, for this procedure to be well-posed,
the solution to these bulk field equations, subject to boundary conditions as
specified above, must be unique. On general grounds we expect that the prescription given here has this property, since we specify enough
data, and we will also illustrate this in the first non-trivial
example below.
\\

\paragraph{Example.} We now illustrate our general discussion in the simplest
possible setup. Namely, we will discuss the duality for the CFT contour of Fig.~\ref{fig:contour}, and compute a two-point function to show that we find the
correct $i\epsilon$ insertions.
A more extended discussion that includes a discussion of examples
corresponding to other time contours (thermal, closed time, etc.), fields, Asymptotically AdS spacetimes etc. will be presented in \cite{SvR}.

\begin{figure}
\includegraphics[width=7cm]{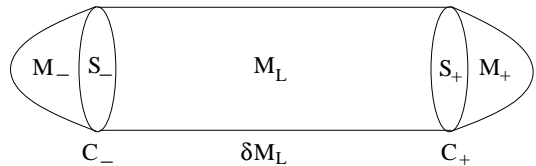}
\caption{\label{fig:cigar} The $CFT_2$ contour, with the  spatial
circle and points at infinity added. Our prescription is to fill it in with
an AdS spacetime consisting of three parts as well.
The corners $C_\pm$ extend to hypersurfaces $S_\pm$ in the bulk, and $\delta M_L$ is the cylindrical conformal boundary of $M_L$.}
\end{figure}

As discussed above, the contour in Fig.~\ref{fig:contour} corresponds to real-time vacuum-to-vacuum correlators. Although the discussion can be easily done for any $CFT_d$ (with a holographic dual), for concreteness we specialize here to $d=2$. Including the spatial $S^1$ direction, we have redrawn
the contour in Fig.~\ref{fig:cigar},
and we have also compactified the Euclidean
semi-infinite cylinders by adding a point at infinity. The corners of the contour are
two circles which we denote as $C_\pm$. The prescription
now amounts to holographically filling in this surface with a bulk
manifold consisting of three components, namely a segment $M_L$ of
Lorentzian AdS$_{3}$ and two `caps' $M_\pm$ consisting of half of
Euclidean AdS$_{3}$. One can view these caps as providing a
Hartle-Hawking wave function on the hypersurfaces $S_\pm$ (where
$\partial S_{\pm} = C_\pm$). In this respect, our prescription
is not only field theory inspired but also in line with standard considerations
on wave functions in quantum gravity \cite{Hartle:1983ai},
see \cite{Maldacena:2001kr,Marolf:2004fy} for a related discussion
in the context of AdS/CFT.

We now propose that the relation between bulk and boundary quantities reads:
\begin{multline}
\label{eq:ansatzrealtimeadscft}
\langle 0| T \exp \Big( - i \int_{\delta M_L} d^d x \sqrt{-g} \phi_{(0)} \op  \Big) |0 \rangle
= \\ \exp\Big(i I_L[\phi_{(0)},\phi_-,\phi_+]
- I_E[0,\phi_-] - I_E[0,\phi_+]\Big).
\end{multline}
with $\delta M_L$ the conformal boundary of $M_L$ as in Fig.~\ref{fig:cigar}, $I_L[\phi_{(0)},\phi_-,\phi_+]$ the \emph{on-shell} Lorentzian
action for $M_L$ that depends not only on $\phi_{(0)}$ but also on initial and final
data $\phi_{\pm}$, and $I_E[\phi_{(0,\pm)}, \phi_\pm]$ the
Euclidean on-shell actions on the half Euclidean spaces $M_{\pm}$
with sources $\phi_{(0,\pm)}$ and boundary condition $\phi_{\pm}$ at $S_\pm$.
In (\ref{eq:ansatzrealtimeadscft}) we set the
sources  $\phi_{(0,\pm)}$ to zero since we are interested in
vacuum-to-vacuum
correlators. Nonzero values for $\phi_{(0,\pm)}$ would correspond to changing the initial and/or final state, as it does in the CFT.
As the notation indicates, we expect (and this
will be verified below) that this procedure leads to time-ordered
products.

Finally, we need to fix $\phi_\pm$ by specifying the behavior of the solution at the corners. This we do by imposing the following two
`matching conditions' for the fields across the $S_\pm$:
\begin{enumerate}
\item As is already indicated in \eqref{eq:ansatzrealtimeadscft}, we impose that the induced values of the bulk fields, so the $\phi_\pm$, are the same on both sides of $S_\pm$.
\item We also demand \emph{stationarity} of the combined on-shell supergravity actions
with respect to variations with respect to $\phi_\pm$:
\begin{equation}
\label{eq:secondmatching}
\frac{\delta}{\delta \phi_\pm} \Big( i I_L[\phi_{(0)},\phi_-,\phi_+]
- I_E[\phi_{(0,-)},\phi_-] - I_E[\phi_{(0,+)},\phi_+] \Big) = 0
\end{equation}
which should be read as an equation for $\phi_\pm$.
\end{enumerate}
Some comments are in order. First, since we think of the bulk
solution as a saddle point to some stringy path integral, these conditions
are a direct consequence of the saddle-point approximation.
Second, taking derivatives of an on-shell action gives the conjugate
momentum, so we obtain the standard junction conditions
\cite{Israel:1966rt}, except for extra factors of $i$. Last, the data $\phi_\pm$ should be compatible with $\phi_{(0)}$ and $\phi_{(0,\pm)}$ at the corners $C_\pm$. In the example below, we will see how this is done.
\\

\paragraph{Two-point function.}
We now specialize to a free massive scalar $\Phi$, propagating
without backreaction on empty AdS$_3$, capped off with two Euclidean
half-balls as in Fig.~\ref{fig:cigar}. Our aim is to
holographically compute the two-point function of the operator
$\mathcal O$ dual to $\Phi$, including the
correct $i\epsilon$-terms, with the above prescription.
The relevant part of the supergravity action is simply:
\begin{equation}
\label{eq:actionscalar}
S = \frac{1}{2}\int d^3 x \sqrt{|G|} (- \partial_\mu \Phi \partial^\mu \Phi - m^2 \Phi^2).
\end{equation}
The dimension of $\mathcal O$ is $\Delta = 1 + \sqrt{1+m^2} = 1
+ l$ with $l \in \{0,1,2, \ldots\}$.

First consider the scalar field solution in the Lorentzian
spacetime without the caps. In the AdS$_3$ background,
\[
ds^2 = - (r^2 + 1) dt^2 + \frac{dr^2}{r^2 + 1} + r^2 d\phi^2\, ,
\]
the mode solutions to the Klein-Gordon equation are of the form
$e^{-i\omega t + ik\phi}f(\omega,\pm k,r)$ with
\begin{multline*}
f(\omega,k,r) =
C_{\omega k l} (1+r^2)^{\omega/2} r^{k}
F(\hat{\omega}_{kl},\hat{\omega}_{kl}-l;k+1;-r^2) \\
= r^{l-1}+ \ldots + r^{-l-1}\alpha(\omega,k,l) [\ln(r^2) + \beta(\omega,k,l)] + \ldots
\end{multline*}
where $\hat{\omega}_{kl} = (\omega + k + 1 + l)/2$, $C_{\omega k l}$
is a normalization factor chosen
such that the coefficient of the leading term equals 1
and in the last line we omitted terms of lower powers of $r$
and some terms polynomial in $\omega$ and $k$ (which would lead to contact terms in the 2-point function). Furthermore,
\begin{eqnarray}
\alpha(\omega,k,l)&=&
(\hat{\omega}_{kl} -l)_l (\hat{\omega}_{kl} -k-l)_l/(l! (l-1)!)\, ,
\nonumber \\
\beta(\omega,k,l) &=& - \psi(\hat{\omega}_{kl}) -
\psi(\hat{\omega}_{kl} -\omega -l)\, ,
\end{eqnarray}
where $(a)_n = \Gamma(a+n)/\Gamma(a)$ is the Pochhammer symbol
and $\psi(x)=d \ln \Gamma(x)/dx$ is the digamma function.
Note also that $f(\omega,k,r) = f(-\omega,k,r)$. Only the $f(\omega,k,r)$ with $k \geq 0$ are regular for $r\to 0$, so the modes we use below are of the form $e^{-i\omega t + ik\phi} f(\omega,|k|,r)$.

We would now like to obtain the most general solution whose leading
asymptotics ($\sim r^{l-1}$ as $r \to \infty$) contain an arbitrary source
$\phi_{(0)}(t,\phi)$ for the dual operator.  This solution is
\begin{widetext}
\begin{equation}
\label{eq:mostgeneralphil}
\Phi(t,\phi,r) = \frac{1}{4\pi^2}\sum_{k \in \mathbb Z} \int_C d\omega
\int d\hat t \int d \hat \phi e^{-i\omega (t - \hat t)
+ ik(\phi - \hat \phi)} \phi_{(0)}(\hat t, \hat \phi) f(\omega,|k|,r)
+ \sum_{\pm} \sum_{k \in \mathbb Z}
\sum_{n = 0}^{\infty} c_{nk}^\pm e^{-i\omega_{nk}^\pm t + ik\phi}g(\omega_{nk},|k|,r)
\end{equation}
\end{widetext}
which we now proceed to explain.
\begin{figure}
\includegraphics[width=7cm]{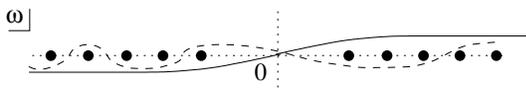}
\caption{\label{fig:freqcontour} Contours around the poles in the complex $\omega$-plane.}
\end{figure}
Clearly, the  first term in (\ref{eq:mostgeneralphil}) has the
correct leading behavior, but we should explain the $C$ representing
a contour in the complex $\omega$-plane. This contour is necessary
to avoid the poles in $\alpha(\omega,k,l) \beta(\omega,k,l)$ at:
\[
\omega = \omega_{nk}^\pm \equiv \pm(2n + k + 1 + l) \, , \quad n \in \{0,1,2,\ldots\}.
\]
We are now completely free to specify any contour that circumvents
the poles, for example the striped contour in
Fig.~\ref{fig:freqcontour}. The difference between two contours is a
sum over the residues:
\begin{eqnarray*}
g(\omega_{nk},k,r) &=& \oint_{\omega_{nk}} d\omega f(\omega_{nk},k,r) \\
&\sim& r^{-l-1} \alpha(\omega_{nk},k,l) \Big( \oint_{\omega_{nk}} d\omega \beta(\omega,k,l) \Big).
\end{eqnarray*}
The $g(\omega_{nk},k,r)$ are the `normalizable modes'.
Since they vanish asymptotically, we
can actually freely add them to the solution $\Phi$ (so not just as
residues) without affecting the fact that $\Phi \sim \phi_{(0)} r^{l-1}$ for
large $r$. Therefore, the most general solution includes a sum over
these normalizable modes with arbitrary coefficients $c_{nk}^\pm$,
as appears in (\ref{eq:mostgeneralphil}). Since a change of
contour can be undone by also changing the $c_{nk}^\pm$, let us fix
the contour to be the solid line in Fig.~\ref{fig:freqcontour}. This
means all the non-uniqueness in the Lorentzian solution is captured
by the $c_{nk}^\pm$.

For later use, let us present an alternative form of the solution.
Without loss of generality, we can assume that the
initial matching surface $S_-$ is at $t=0$ and that the sources are
zero in the vicinity of $S_-$. Then, near $S_-$, we can perform the
$\omega$-integral by closing the contour and picking up the poles in
$f(\omega,k,r)$, resulting in
\begin{multline*}
\Phi = \frac{1}{4\pi^2}\sum_{n=0}^\infty
\sum_{k \in \mathbb Z} e^{-i \omega_{nk}^- t  + ik \phi}
\phi_{(0)} (\omega_{nk}^-,k) g(\omega_{nk},|k|,r) \\
+ \sum_{\pm} \sum_{k \in \mathbb Z}
\sum_{n = 0}^{\infty} c_{nk}^\pm
e^{-i\omega_{nk}^\pm t + ik\phi}g(\omega_{nk},|k|,r)
\end{multline*}
where we Fourier transformed the source. Of course, this is
an expected result; it just represents the
completeness of the modes.

Now consider the solution on the `initial cap', so on the space
specified by the metric,
\[
ds^2 = (r^2 + 1) d\tau^2  + \frac{dr^2}{r^2 +1} + r^2 d\phi^2
\]
with $-\infty < \tau \leq 0$, so that we have half of
Euclidean AdS space. Had the bulk been the entire Euclidean AdS space,
the Klein-Gordon equation would have a unique regular solution
given boundary data. In particular, with zero sources the unique regular
solution is identically  equal to zero. In our case the sources are zero
but we only consider half of the space, so solutions that would be excluded
are now allowed because they are only singular at the other half of
the space. These regular solutions are precisely the analytically continued
Lorentzian normalizable modes, so we find solutions when $\omega =
\omega_{nk}^\pm$. Since the solution should vanish at $\tau
\to - \infty$,
the most general Euclidean solution contains only negative frequencies,
\[
\Phi(\tau,\phi,r) = \sum_{n,k} d_{nk}^-
e^{-\omega_{nk}^- \tau + ik\phi}g(\omega_{nk},|k|,r)\, ,
\]
with thus far arbitrary coefficients $d_{nk}^-$.

We can now consider the matching at
$\tau = t = 0$, which will fix the initial data. From the continuity
$\Phi_L(0,\phi,r) = \Phi_E(0,\phi,r)$ we find, using orthogonality
and completeness of the $g(\omega_{nk},|k|,r)$:
\[
\phi_{(0)}(\omega_{nk}^-,k) + c_{nk}^- + c_{nk}^+ = d_{nk}^-
\]
Eqn. (\ref{eq:secondmatching}) yields a relation between conjugate momenta,
\[
- i \partial_t \Phi_L  = \partial_\tau \Phi_E\, .
\]
Substituting the solutions we find
\[
- \omega_{nk}^- \phi_{(0)}(\omega_{nk}^-,k) -
\omega_{nk}^- c_{nk}^- -\omega_{nk}^+ c_{nk}^+ = - \omega_{nk}^- d_{nk}^-\, ,
\]
so that $c_{nk}^+ = 0$. Similarly, the matching to the out state
determines $c_{nk}^- = 0$, and indeed all the freedom in the
bulk solution is fixed. We remark that, had we chosen any other
contour in (\ref{eq:mostgeneralphil}), we would have found
nonzero values of some of the $c_{nk}^\pm$, effectively throwing us
back to the solid line of Fig.~\ref{fig:freqcontour}.

Finally, the two-point function is obtained from the
$r^{-l-1}$ term in the asymptotic expansion of
(\ref{eq:mostgeneralphil}) (with  $c_{nk}^\pm = 0$):
\begin{multline*}
\langle 0| T \op(t,\phi) \op(0,0)  | 0 \rangle=\\
\frac{l}{4\pi^2 i} \sum_k \int_C d\omega e^{-i\omega t + ik\phi} \alpha(\omega,|k|,l)
\beta(\omega,|k|,l).
\end{multline*}
with the contour $C$ being the same as for the bulk solution, thus
the standard Feymnan prescription leading to time ordered correlators.
We emphasize again that $C$ was completely fixed by the matching to the
caps. Integrating over $C$ is equivalent to integrating over the real axis 
and shifting $\omega \to \omega(1+ i \epsilon)$. The Fourier transform of this 
expression then gives
\[
\langle 0| T \op(t,\phi) \op(0,0) |0 \rangle
= \frac{l^2/(2^{l+1} \pi)}{[\cos(t -i\epsilon t) - \cos(\phi)]^{l+1}}\, .
\]
This is the expected form for a time-ordered
two-point function on a cylinder and the normalization coefficient 
can be shown to agree with the standard AdS/CFT normalization of
2-point functions.
\\

\paragraph{Conclusion.}
We presented a prescription that relates in- and out-states of the
boundary QFT to initial and final data for the bulk fields.
We discussed in detail the case of a free bulk scalar field in pure AdS,
but the procedure extends to other contours, fields (including the metric), asymptotically AdS spacetimes and higher $n$-point functions in a clear manner, details of which will be presented in \cite{SvR}.
The prescription allows us to study holographically QFT dynamics in cases
where analytic continuation from the Euclidean regime
does not suffice. It also offers a
new perspective on the holographic encoding of bulk spacetimes, since
the state or density matrix corresponding to a given geometry is directly
related to the Euclidean parts of the solution.
This may allow us to understand how regions
beyond bulk horizons are `encoded' in the QFT data. We hope to address this
and other intriguing aspects of real-time gauge/gravity duality in the
near future.

\end{document}